\documentclass[12pt]{article}
\usepackage{enumerate}
\usepackage{ifpdf}
\usepackage{ae}
\usepackage[T1]{fontenc}
\usepackage[ansinew]{inputenc}
\usepackage{amsmath}
\usepackage{amssymb}
\usepackage{graphicx}
\usepackage{color}
\definecolor{darkblue}{cmyk}{0.9,0.9,0,0}
\definecolor{darkgreen}{rgb}{0,0.55,0}
\usepackage[colorlinks=true,linkcolor=darkblue,citecolor=darkblue,urlcolor=darkblue]{hyperref}
\usepackage{epsfig}
\usepackage{epstopdf}

\makeatletter
\long\def\@makecaption#1#2{
  \vskip\abovecaptionskip
  \sbox\@tempboxa{{\captionfonts #1: #2}}
  \ifdim \wd\@tempboxa >\hsize
    {\captionfonts #1: #2\par}
  \else
    \hbox to\hsize{\hfil\box\@tempboxa\hfil}
  \fi
  \vskip\belowcaptionskip}
\makeatother

\newcommand{\beq}{\begin{equation}}
\newcommand{\eeq}{\end{equation}}
\newcommand{\beqy} {\begin{eqnarray}}
\newcommand{\eeqy} {\end{eqnarray}}
\newcommand{\bsmat}{\begin{smallmatrix}}
\newcommand{\esmat}{\end{smallmatrix}}
\newcommand{\bmat}{\begin{matrix}}
\newcommand{\emat}{\end{matrix}}

\def\({\left(}
\def\){\right)}
\def\[{\left[}
\def\]{\right]}

\def\<{\langle}
\def\>{\rangle}

\usepackage{varioref}
\usepackage{makeidx}
\makeindex

\usepackage[english]{babel}
\begin{document}

\thispagestyle{empty}

\renewcommand{\thefootnote}{\fnsymbol{footnote}}
\setcounter{page}{1}
\setcounter{footnote}{0}
\setcounter{figure}{0}
\begin{center}
$$$$

{\LARGE\textbf{\mathversion{bold}
Higher-spin correlators}}
\vspace{1.0cm}

\textrm{\Large Luis F. Alday and Agnese Bissi}
\\ \vspace{1.2cm}

\textit{Mathematical Institute, University of Oxford,}  \\
\textit{24-29 St Giles', Oxford, OX1 3LB, UK} \\
\vspace{5mm}

\par\vspace{1.5cm}

\textbf{Abstract}\vspace{2mm}
\end{center}

\noindent
We analyze the properly normalized three-point correlator of two protected scalar operators and one higher spin twist-two operator in ${\cal N}=4$ super Yang-Mills, in the limit of large spin $j$. The relevant structure constant can be extracted from the OPE of the four-point correlator of protected scalar operators. We show that crossing symmetry of the four point correlator plus a judicious guess for the perturbative structure of the three-point correlator, allow to make a prediction for the structure constant at all loops in perturbation theory, up to terms that remain finite as the spin becomes large. Furthermore, the expression for the structure constant allows to propose an expression for the all loops four-point correlator ${\cal G}(u,v)$, in the limit $u,v \rightarrow 0$. Our predictions are in perfect agreement with the large $j$ expansion of results available in the literature.

\vspace*{\fill}

\setcounter{page}{1}
\renewcommand{\thefootnote}{\arabic{footnote}}
\setcounter{footnote}{0}

\newpage

 \def\nref#1{{(\ref{#1})}}



\section{Introduction}

${\cal N}=4$ super Yang-Mills is a CFT and as such should be characterized by the spectrum of anomalous dimensions $\Delta_i$ of its local gauge invariant operators ${\cal O}_i$, or two-point functions, and its structure constants $C_{ijk}$, or three-point functions. These quantities are usually complicated functions of the coupling constant $g$ and the number of colors $N$, but a vast simplification occurs in the planar limit, in which they are non-trivial functions of $a=\frac{g^2 N}{4\pi^2}$. Over the last few years, there has been an enormous success in computing the anomalous dimensions $\Delta_i(a)$ at all values of the coupling constant for large class of operators, by using the methods of integrability, see for instance \cite{Beisert:2010jr}.

Two important classes of operators are protected operators and higher-spin operators. Anomalous dimensions of protected operators do not get quantum corrections, and hence they are equal to the bare dimension at all values of the coupling $\Delta(a) =\Delta_0$. Likewise, non-renormalization theorems protect the structure constants of protected operators from quantum corrections \cite{Lee:1998bxa}. The anomalous dimension of higher-spin operators $\Delta_j(a)$ is more interesting. For instance, for large values of the Lorentz spin $j$, it grows logarithmically \cite{largej}:
\begin{equation}
\label{cusp}
\gamma_j(a) \equiv \Delta_j(a) -j -2= f(a) (\log j+\gamma_e) + g(a) + {\cal O}(1/j)  
\end{equation}

\noindent where $f(a)$ is the so-called cusp or universal anomalous dimension, while $g(a)$ depends on the details of the operator under consideration. The cusp anomalous dimension appears in several other computations ({\cal e.g.} it controls IR singularities of scattering amplitudes as well as UV singularities of Wilson loops with cusps), which makes the study of higher spin operators even more interesting.  Over the last few years, there has been remarkable progress in the computation of the cusp anomalous dimension for planar ${\cal N}=4$ super Yang-Mills, to the point that it is now known to all values of the coupling \cite{Beisert:2006ez,Basso:2007wd}. On the other hand, much less is known about three-point functions involving higher-spin operators. Some specific examples at one loop were first discussed in \cite{Alday:2005nd}. More recently, \cite{Kazakov:2012ar}  computed the structure constants of three higher-spin operators at tree-level \footnote{See \cite{Georgiou:2011qk,Kazama:2011cp} for related computations involving higher-spin operators}. The focus of this paper will be the properly normalized structure constant involving two protected operators and one higher-spin operator, sometimes referred to as universal structure constant and schematically of the form
\begin{equation}
C_j (a)\sim \frac{\langle  {\cal O} {\cal O} {\cal O}_j  \rangle}{ \langle  {\cal O} {\cal O} \rangle \sqrt{\langle  {\cal O}_j  \tilde{\cal O}_j\rangle}  }
\end{equation}
where ${\cal O}$ is a protected scalar operator and  ${\cal O}_j$ is a higher-spin operator with Lorentz spin $j$. There are two distinct ways to compute this quantity. The first one is by doing an explicit/Feynman diagram computation. This was done at one loop in \cite{Plefka:2012rd}. The second way is by starting with the four point correlation function of protected scalar operators $\langle {\cal O}_1 {\cal O}_2  {\cal O}_3 {\cal O}_4 \rangle $ and then considering a coincidence limit in which ${\cal O}_2$ approaches ${\cal O}_1$ and ${\cal O}_4$ approaches ${\cal O}_3$. In this limit we have the usual OPE expansion $ {\cal O}_1 {\cal O}_2 \sim \sum_j C_j {\cal O}_j$ and hence $C_j$ can be extracted from the four point function
\begin{equation}
\langle {\cal O}_1 {\cal O}_2  {\cal O}_3 {\cal O}_4 \rangle  \sim \sum_j C_j C_j
\end{equation}
This OPE was analyzed in full detail in \cite{Dolan:2004iy}. The one loop result can be extracted from this analysis and is in perfect agreement with  the result of \cite{Plefka:2012rd}. Finally the OPE, together with explicit results at three-loops for the four point function \cite{Eden:2011we}, was used by \cite{Eden:2012rr} to extract the universal structure constants at three-loops. 

The aim of the present paper is to understand the large $j$ behavior of the universal structure constant $C_j(a)$. Based on the OPE structure, symmetry arguments and intuition from the available perturbative results we are able to guess an expression analogous to (\ref{cusp}). More precisely, we find that, in the large $j$ limit and at all orders in perturbation theory:
\begin{equation}
\frac{C_j(a)}{C_j^{tree}} = F(a)2^{-  \gamma_j(a)/2 } e^{- \gamma_e  \gamma_j(a)/2} e^{-  g(a)/2( \log j+\gamma_e)} \Gamma \left( 1- \frac{ \gamma_j(a) }{2} \right)+{\cal O}(1/j)
\end{equation}
where $F(a)$ is a function of the coupling constant, independent of $j$ and $\gamma_j(a)$ is the anomalous dimension of the higher spin operator, which behaves according to (\ref{cusp}) in the large spin limit. 

Having a proposal for the structure constant in the large $j$ limit, it turns out we can derive an expression for the $u,v \rightarrow 0$ limit of the four-point correlator

\begin{equation}
\lim_{u,v \rightarrow 0}{\cal G}(u,v) \sim e^{-\frac{f(a)}{4} \log u \log v +\frac{g(a)}{2}  (\log u + \log v)} J(u, v)
\end{equation}
where $J(u, v)$ is a symmetric function of $u,v$ and is given to all loops. The organization of this paper is as follows: in the next section we introduce the precise correlator to be studied and review its computation to one loop. The computation can be done in two ways: explicit Feynman diagram calculation and from the OPE of a specific four-point correlator. The later method is more appropriate to understand the systematics of a large spin expansion. In section three we analyze the correlator in the high spin limit and propose an all loop expression that fully captures this expansion. Furthermore, we use our proposal in order to derive an all loop expression for the four-point correlator mentioned above, in the limit of vanishing cross-ratios.  We end up with some conclusions and open problems. Finally, in the appendix we compute the large spin expansion of the results available in the literature and show that they are in full agreement with our prediction. 

\section{The universal structure constant}

Let us start by reviewing the explicit computation of \cite{Plefka:2012rd}. We want to study the three-point function involving two protected scalar operators
\begin{eqnarray}
{\cal O}(x) = Tr \left( \bar \Phi_{12}  \bar \Phi_{13} \right),~~~ \tilde {\cal O}(x) = Tr \left( \bar \Phi_{12}  \Phi^{13} \right)
\end{eqnarray}
and one twist-two higher-spin operator, which at tree-level is of the form \cite{Ohrndorf:1981qv}
\begin{eqnarray}
{\cal O}_j(x) =z^{\mu_1}...z^{\mu_j} \sum_{k=0}^j(-1)^k \left( \begin{matrix} j \\ k \end{matrix}\right)^2Tr \left( D_{\mu_1}... D_{\mu_k} \Phi^{12} D_{\mu_{k+1}}...D_{\mu_j} \Phi^{12} \right)
\end{eqnarray}
where $z^\mu$ is a null vector. The structure of the two and three-point functions under consideration is fixed by conformal invariance
\begin{eqnarray}
\langle {\cal O}_j(x_1) {\cal O}_k(x_2)  \rangle &=& \delta_{jk} {\cal N}_j(a) \frac{(z \cdot x_{12} )^{2j}}{|x_{12}^2|^{2j+2+\gamma_j(a)}} \\
 \langle {\cal O}(x_1)\tilde {\cal O}(x_2) {\cal O}_j(x_3) \rangle &=&  C_{{\cal O O} j}(a) \frac{\left( \frac{1}{2}z_\mu  \partial^\mu_{x_3} \log \left(\frac{x_{23}^2}{x_{13}^2} \right)\right)^j}{|x_{12}|^{2-\gamma_j(a)}|x_{13}|^{2+\gamma_j(a)}|x_{23}|^{2+\gamma_j(a)}} 
\end{eqnarray}
where we have assumed $j$ even. $ \cite{Plefka:2012rd}$ explicitly computed ${\cal N}_j(a)$ and $C_{{\cal O O} j}(a)$ up to one loop, obtaining the following result for the structure constant
\begin{equation}
\label{plefka}
C_j(a) = \frac{C_{{\cal O O} j}(a)}{{\cal N}_0 {\cal N}_j(a)^{1/2}} = 2^{1/2} \frac{\Gamma(j+1)}{\Gamma(2j+1)^{1/2}}\left(1+ \left( 2 S_1(j) (S_1(j) - S_1(2j) ) -S_2(j) \right) \frac{a}{2}+... \right)
\end{equation}
where ${\cal N}_0=\frac{1}{(8\pi^2)^2}$ comes form the normalization of the protected operators and the Harmonic sums are given by $S_a(j)= \sum_{n=1}^j \frac{1}{n^a}$.

\subsection{Extracting the universal structure constants from the OPE}

The above result can also be obtained by an alternative method, see e.g. \cite{Dolan:2004iy}, which we briefly describe in the following. We start with the four-point function of protected scalar operators located at $x_{1,2,3,4}$. Conformal invariance implies the following structure

\begin{eqnarray}
\langle {\cal O}_1(x_1)  {\cal O}_2(x_2)  {\cal O}_3(x_3)  {\cal O}_4(x_4) \rangle = \frac{{\cal G}(u,v)}{x_{12}^2 x_{23}^2 x_{34}^2 x_{41}^2}  
\end{eqnarray}
so that this correlator can depend non-trivially only on the cross-ratios

\begin{equation}
u = \frac{x_{12}^2 x_{34}^2}{x_{13}^2 x_{24}^2},~~~~~ v = \frac{x_{14}^2 x_{23}^2}{x_{13}^2 x_{24}^2}
\end{equation}
next we consider the OPE limit $x_{12} \rightarrow 0$ and $x_{34} \rightarrow 0$. At the level of the cross-ratios this means $u \rightarrow 0$ and $v \rightarrow 1$ (but with $u$ approaching zero faster). When taking the OPE limit, the operators ${\cal O}_1$ and ${\cal O}_2$ will give rise to an infinite sum of conformal primaries and their descendants

\begin{equation}
{\cal O}_1(x_1) {\cal O}_2(x_2) \sim \sum_{\ell}C_{12,\ell}{ \frac{x_{12}^{\mu_1} ... x_{12}^{\mu_j}}{x_{12}^{\Delta_\ell}}} {\cal O}_\ell^{\mu_1 ... \mu_j}(x_1)
\end{equation}
A conformal primary is labelled by $\ell$, which encodes the $SU(4)$ representation, the dimension and the Lorentz spin $j$. Given a primary all its descendants are packed into a conformal partial wave. Hence, in this limit we can write the four point function as a sum over intermediate states, propagating from the left side to the right side. Each term is weighted by $C_{12,\ell} C_{34,\ell}$, the probability of creating the state on the left and reabsorbing it on the right.  In order to compute the structure constant we are interested in, we need to select the appropriate primaries. The protected scalar primaries we started with transform in the ${\bf 20}$ of $SU(4)$. The states on the OPE will transform under the ${\bf 20} \times {\bf 20} = {\bf 1}+{\bf 15}+{\bf 20}+...$, so first we need to project over the ${\bf 20}$. Furthermore, as we are interested only in twist two operators, we need to take the limit $u \rightarrow 0$. Higher powers of $u$ will correspond to higher twist and can be dropped. Finally, the power of $Y=1-v$ will characterize the spin of the exchanged operator.  See \cite{Dolan:2004iy,Eden:2012rr} for more details. 

Let us show how to re-obtain the result of the previous subsection. The equation to be solved is the following  \cite{Dolan:2004iy}

\begin{eqnarray}
\label{wave}
\left(\frac{1}{1-Y}+1\right) -\frac{Y}{1-Y} \left( \frac{1}{2} \log(1-Y)\log u+Li_2(Y) \right) a+...=\\
= \sum_{j=0}^\infty  \left(C_j(a)\right)^2 u^{\gamma_j(a)/2} Y^j ~_2F_1(j+1+\gamma_j(a)/2,j+1+\gamma_j(a)/2,2j+2+\gamma_j(a);Y) \nonumber
\end{eqnarray}
The left hand side of this equation is the corresponding limit of the four-point correlation function up to one loop, which is explicitly known, projected over intermediate states in the ${\bf 20}$ representation. As can be seen, all powers of $u$ have been dropped. The right hand side represents the sum over conformal waves corresponding to twist-two primaries of spin $j$. $\gamma_j(a)$ is the anomalous dimension of the corresponding operator and $C_j(a)$ the properly normalized structure constant. The anomalous dimension and structure constant admit a perturbative expansion
\begin{eqnarray}
\gamma_j(a)&=&\gamma_1 a + ...\\
C_j(a) &= &C_j^{tree} + a C_j^{(1)}+...
\end{eqnarray}
The task is then to solve for $\gamma_j(a)$ and $C_j(a)$ order by order in perturbation theory and for each value of $j$. This can be done by expanding (\ref{wave}) in powers of $Y$ and then equating the two sides term by term. The solution is
\begin{eqnarray}
C_j^{tree} &=&2^{1/2} \frac{\Gamma(j+1)}{\Gamma(2j+1)^{1/2}},\\
C_j^{(1)} &=& 2^{-1/2} \frac{\Gamma(j+1)}{\Gamma(2j+1)^{1/2}} \left( 2 S_1(j) (S_1(j) - S_1(2j) ) -S_2(j) \right)\\
\gamma_1 &= &2 S_1(j)
\end{eqnarray}
in perfect agreement with the explicit computation (\ref{plefka}). Note that this method provides not only the structure constant, but also the anomalous dimension (as does the explicit computation), but one does not need to specify the precise quantum state that is propagating. Furthermore, note that the OPE analysis gives directly the properly normalized structure constants. Finally, let us add that this procedure has been carried out up to three loops in \cite{Eden:2012rr}. The corresponding $C_j(a)$ is presented in the appendix.

\section{Large spin analysis}

We want to understand the large $j$ behavior of the universal structure constants $C_j(a)$, as extracted from the OPE of the four-point function of scalar protected operators. As reviewed in the previous section, the equation to be solved is the following

\begin{equation}
\label{OPE}
\left(\frac{1}{1-Y}+1\right) + ... = \sum_{j=0}^\infty 2\frac{\Gamma(j+1)^2}{\Gamma(2j+1)}  \hat C_j(a) u^{\gamma_j(a)/2} Y^j ~_2F_1(j+1+\gamma_j(a)/2,j+1+\gamma_j(a)/2,2j+2+\gamma_j(a);Y)
\end{equation}
where we have defined
\begin{equation}
 \hat C(j) = \left(\frac{C_j(a)}{C_j^{tree}} \right)^2
\end{equation}
Let us focus in the r.h.s. of (\ref{OPE}) for large values of $j$. As $j$ increases, most of the contribution comes from the region $Y \approx 1$. The appropriate limit to consider is
\begin{eqnarray}
Y=1-v,~~~~~j= \frac{x}{\sqrt{v}}
\end{eqnarray}
and then take $v \rightarrow 0$. In this limit, the sum over $j$ becomes an integral with the natural measure $\sum_j \rightarrow \frac{1}{2} \int_0^\infty \frac{dx}{\sqrt{v}}$ \footnote{The factor of $\frac{1}{2}$ comes from the fact that the sum over $j$ is only for even spin}. Furthermore, in this limit \footnote{The large $j$ limit of the conformal block $I_{j}(Y) = \frac{\Gamma(j+1)^2}{\Gamma(2j+1)}  Y^j ~_2F_1(j+1+\gamma/2,j+1+\gamma/2,2j+2+\gamma;Y)$ can be understood as follows, see for instance, \cite{Komargodski:2012ek}. Using the integral representation for the hypergeometric function we can write
$$I_{j}(Y)  = \frac{\Gamma(j+1)^2}{\Gamma(2j+1)}  \frac{\Gamma(2j+2+\gamma)}{\Gamma(j+1+\gamma/2)^2} Y^j \int_0^1 t^{j+\gamma/2}(1-t)^{j+\gamma/2}(1-t Y)^{-(j+1+\gamma/2)}dt$$
setting $Y=1-v,~j= \frac{x}{\sqrt{v}}$ and taking the $v \rightarrow 0$ limit, the prefactor turns into $\frac{2^{1+\gamma} x}{\sqrt{v}}$. On the other hand, upon changing coordinates $t \rightarrow 1- t \sqrt{v}$, in the limit the integral reduces to
$$\int_0^\infty \frac{e^{-(t+1/t)x}}{t}=2 K_0(2x)$$
giving the desired result.
}

\begin{equation}
\frac{\Gamma(j+1)^2}{\Gamma(2j+1)}  Y^j ~_2F_1(j+1+\gamma/2,j+1+\gamma/2,2j+2+\gamma;Y) \rightarrow \frac{1}{\sqrt{v}} 2^\gamma 4 x K_0(2x)
\end{equation}

This is a generalization of the limit studied in \cite{Alday:2010zy}. Hence, as $v$ goes to zero, (\ref{OPE})  becomes 

\begin{equation}
\frac{1}{v} (1+2  \sum_{\ell=1} a^\ell F^{(\ell)}|_{u,v=0}) = \frac{1}{v} \int_0^\infty dx 4 \hat C(\frac{x}{\sqrt{v}}) u^{\gamma/2} 2^\gamma x K_0(2x)
\end{equation}
On the left-hand-side we have shown explicitly the leading power in $v$. $F^{(\ell)}|_{u,v=0}$ is the contribution at $\ell-$loops, where we suppress powers of $u$ and $v$, but $\log$'s are in general present. Here comes an important point: the left hand side contribution $ (1+2  \sum_{\ell=1} a^\ell F^{(\ell)}|_{u,v=0})$, even though not known explicitly beyond three-loops, has to be a symmetric function of $u$ and $v$. This is a consequence of crossing-symmetry for the four-point correlator, which is recovered in the limit under consideration. On the right hand side we have suppressed the coupling constant dependence but both, $\gamma$ and $\hat C$, depend on the coupling constant. Hence we arrive at the following equation

\begin{equation}
\label{crossing}
 \int_0^\infty dx 4 \hat C(\frac{x}{\sqrt{v}}) (4 u)^{( f(a) (\log \frac{x}{\sqrt{v}} +\gamma_e) + g(a))/2} x K_0(2x) =L(u,v)
\end{equation}
where $L(u,v)$ is the $u,v \rightarrow 0$ limit of the four-point function ${\cal G}(u,v)$, and is a symmetric function under the interchange of $u$ and $v$. 

What can we say about $\hat C(\frac{x}{\sqrt{v}})$? The large  $j$ expansion of the explicit results (see appendix) suggests the following structure at all loops
\begin{equation}
\log \hat C(j) =  \left( b_{10}  \log j' +b_{11}  \right) a+ \left( b_{20} \log^2 j' +b_{21} \log j'+b_{22}\right)a^2 +   \left( b_{30} \log^3 j'+ ... +b_{33}\right)a^3+...
\end{equation}
where $j'=j e^{\gamma_e}$. This general expansion can be plugged into (\ref{crossing}) and we can integrate order by order in perturbation theory. All integrals are of the form $\int_0^\infty 4 x \log^q x K_0(2x)$ and can be easily performed with Mathematica. Quite remarkably, just by requiring the final result to be symmetric under interchange of $u$ and $v$, we can fix all the coefficients $b_{ij}$ for $i<j$ in terms of the anomalous dimension functions $f(a)$ and $g(a)$. We have checked that this is the case up to several loops. For instance, for the first coefficients we find
\begin{eqnarray}
b_{10} = -f_1 \log 2,~~~b_{20}= \frac{1}{2}\zeta_2 f_1^2,~~~b_{21} = \frac{1}{2}\zeta_2 f_1 g_1 - f_2 \log 2-g_2
\end{eqnarray}
and so on, where $f_\ell$ and $g_\ell$ are the $\ell-$loop contributions to $f(a)$ and $g(a)$. This can be seen to agree precisely with the large $j$ expansion of the results available in the literature, shown in the appendix. Furthermore, the structure of the leading coefficients suggests the following closed form
\begin{equation}
\hat C(j) = 2^{-  \gamma_j(a) } e^{- \gamma_e  \gamma_j(a)} e^{-  g(a)( \log j+\gamma_e) } \left[ \Gamma \left( 1- \frac{ \gamma_j(a) }{2} \right) \right]^2
\end{equation}
which indeed captures all the logarithmic terms (not only the leading ones) to the order we checked!\footnote{This is to be contrasted with the simple scaling behavior of the DIS Wilson loop coefficient, see {\it e.g.} \cite{Bianchi:2013sta}. However, in both quantities the large $j$ behavior is controlled by the anomalous dimension, as expected \cite{Korchemsky:1993uz}} We will see in the next section that indeed this leads to a function symmetric under interchange of $u$ and $v$. This expression for $\hat C(j) $ implies the large spin expansion for the universal structure constant cited in the introduction, namely 

\begin{equation}
\label{Cproposal}
\frac{C_j(a)}{C_j^{tree}} = F(a)2^{-  \gamma_j(a)/2 } e^{- \gamma_e  \gamma_j(a)/2} e^{-  g(a)/2( \log j+\gamma_e)} \Gamma \left( 1- \frac{ \gamma_j(a) }{2} \right)+{\cal O}(1/j)
\end{equation}
Recall that, at this order, $ \gamma_j(a)$ behaves as in (\ref{cusp}). This determines the large $j$ expansion of the structure constant, to all loops in perturbation theory and is the main result of this paper. It is interesting to note that we have never used the explicit expression for $L(u,v)$  in (\ref{crossing}), just its symmetry properties. In spirit, this is very similar to the conformal bootstrap.

In the appendix we compute the large $j$ expansion of the structure constant from the available results in the literature (up to three loops) and check  precise agreement with our prediction. Furthermore, in doing the comparison we can compute

\begin{equation}
F(a) = -\frac{1}{2} \zeta_2 a+ \frac{7}{8} \zeta_2^2 a^2 + ...
\end{equation}

\subsection{Implications for the four-point function}

Having a prediction for the large spin behavior of $\hat C(j)$ at all orders in perturbation theory, we can ask what are the implications for the original four-point function in the limit $u,v \rightarrow 0$

\begin{equation}
\lim_{u,v \rightarrow 0} {\cal G}(u,v) \equiv L(u,v)
\end{equation}
 In order to answer this we simply insert (\ref{Cproposal}) in (\ref{crossing}) and compute the resulting $L(u,v)$. By using the following identity

\begin{equation}
\label{Kidentity}
\int_0^\infty dy\, 4y \, y^k \, K_0(2y) =  \left[ \Gamma \left( 1+ \frac{ k}{2} \right) \right]^2
\end{equation}
we can write (\ref{crossing}) in a symmetric fashion

\begin{eqnarray}
\label{fourlimit}
L(u,v) &=& fin \times e^{-\frac{f(a)}{4} \log u \log v +\frac{g(a)}{2}  (\log u + \log v)} J(\hat u, \hat v)\\
J(\hat u, \hat v) &= & e^{\gamma_e(\hat u +\hat v)}16 \int_0^\infty dx dy \,x y \,x^{\hat u }\, y^{\hat v} \,e^{-f(a) \log x \log y } K_0(2x) K_0(2y) \nonumber \\
\end{eqnarray}
where $fin$ is independent of $u$ and $v$, and not fixed by our analysis. We have introduced the notation 
\begin{eqnarray*}
\hat u = f(a)/2 \log u -\gamma_e f(a) -g(a) \\
\hat v = f(a)/2 \log v -\gamma_e f(a) -g(a) 
\end{eqnarray*}
This shows that for our choice of $\hat C(j)$, the final integral is indeed a symmetric function! $J(\hat u,\hat v)$ is a sum of factorized contributions of the form

\begin{equation}
J(\hat u,\hat v) = e^{\gamma_e(\hat u +\hat v)} \sum_\ell \frac{(-1)^\ell f(a)^\ell}{\ell!} {\cal J}^{(\ell)}(\hat u) {\cal J}^{(\ell)}(\hat v)
\end{equation}
the integrals ${\cal J}^{(\ell)}(k)$ can easily be calculated starting from (\ref{Kidentity}) and taking derivatives with respect to $k$ on both sides. For the first few cases we obtain

\begin{eqnarray}
{\cal J}^{0}(k) &=&  \left[ \Gamma \left( 1+ \frac{ k}{2} \right) \right]^2\\
{\cal J}^{1}(k) &=& \frac{d {\cal J}^{0}(k)}{dk} = {\cal J}^{0}(k) \psi^{(0)}(1+\frac{k}{2})\\
{\cal J}^{2}(k) &=& \frac{d^2 {\cal J}^{0}(k)}{dk^2} = {\cal J}^{0}(k) \left( \psi^{(0)}(1+\frac{k}{2})^2+\frac{1}{2} \psi^{(1)}(1+\frac{k}{2}) \right)
\end{eqnarray}
and so on. The full result can be formally written as 

\begin{equation}
J(\hat u,\hat v) = e^{\gamma_e(\hat u +\hat v)}  e^{-f(a) \frac{\partial^2}{\partial \hat{u} \partial \hat{v}}}  \left[ \Gamma \left( 1+ \frac{\hat u}{2} \right) \right]^2  \left[ \Gamma \left( 1+ \frac{ \hat v}{2} \right) \right]^2
\end{equation}

The limit $u,v \rightarrow 0$  of the four point correlator, up to two loops, was also studied in \cite{Alday:2010zy}, where the sequential null limit of the insertion points was considered. Our results should be compared to section four of that paper \footnote{The relation between the cross-ratios used in \cite{Alday:2010zy} and here is $z \bar{z} = v$ and $(1-z)(1-\bar{z})=u$.}: we see precise agreement between the exponential terms in (\ref{fourlimit}) and the first line of eq. (4.36) in that paper. Furthermore, the first term for $J(\hat u,\hat v)$ precisely agrees with $J$ up to two loops, as computed in  \cite{Alday:2010zy}! Note that our extra factor of $e^{\gamma_e(\hat u +\hat v)}$ will exactly cancel $\gamma_e$ terms in a perturbative expansion, so $J$ will start contributing at two loops. In this paper we have given $J$ at all orders in perturbation theory. 

\section{Conclusions}

We have analyzed the properly normalized three-point correlator of two protected scalar operators and one higher spin operator, in the limit of large spin $j$. The relevant structure constant can be extracted from the OPE of the four-point correlator of protected scalar operators. We have shown that crossing symmetry of the four point correlator, plus a judicious guess for the perturbative structure of the three-point correlation, allow us to make a prediction for the structure constant at all loops in perturbation theory, up to terms that are finite as the spin becomes large. This prediction is in perfect agreement with the available results (namely up to three loops). Furthermore, the expression for the structure constant allows us to propose an expression for the all loops four-point correlator ${\cal G}(u,v)$, in the limit $u,v \rightarrow 0$. Again, this expression is in perfect agreement with the available expressions in the literature (namely up to two loops). There are several open problems

\begin{itemize}

\item Can we extrapolate our results to strong coupling? We have presented results to all loops, for a non-protected three-point correlator,  but still in a perturbative expansion, so that $\gamma_{j}(a)$ is small. The extrapolation may be subtle due to order of limits issues and in particular note that (\ref{Cproposal}) contains poles at intermediate values $\gamma_j(a)=2,4,...$. Note that at these values, twist-two contributions can mix with higher twists. 

\item Once the extrapolation is understood, can we match the result by using $AdS/CFT$? At strong coupling it is not known yet how to compute the correlator of two light operators with a heavy one. On the other hand there are available results for the four-point correlator \cite{Arutyunov:2000ku}, so a comparison may in principle be possible. 

\item Over the last couple of years integrability techniques have been applied to the problem of correlators in ${\cal N}=4$ SYM, see for instance \cite{Gromov:2012vu} and references therein. Can the results of this paper be used to extend this program? certainly higher-spin operators played a very important role in applying integrability to the spectral problem, so we expect the same to happen for correlators. 

\item Similar three-point correlation functions (but transforming as a singlet under $SU(4)$) were considered in \cite{Costa:2012cb}, where Regge theory was applied to correlation functions. It would be interesting to study the large spin behavior of such correlators. Conversely, one could study the structure of poles in the correlators studied in this paper (when analitically continued in $j$). This would be very interesting, as one would be able to make all loop predictions in different regimes.

\item In this paper we gave an expression for the four-point correlator ${\cal G}(u,v)$, in the limit $u,v \rightarrow 0$. Note that this didn't require any explicit computation, but only understanding the structure of the OPE and crossing-symmetry. Can we extend this analysis to general cross-ratios? or equivalently, can our results be used to constraint the four-point correlator at higher loops? in spirit our method is very similar to the conformal bootstrap. It would be interesting to find a relation to recent developments in this area, see {\it e.g.} \cite{Costa:2011dw,Komargodski:2012ek,Fitzpatrick:2012yx}.
\item Related to above point, it would be interesting to understand the results of this paper by using the picture of \cite{Alday:2010zy,Alday:2007mf}.
\item Finally, note that we didn't use much of the structure of ${\cal N}=4$ SYM, so we expect our results to be valid for other CFT with certain modifications. It would be interesting to study this problem in detail.
\end{itemize}

\subsection*{Acknowledgments}
\noindent
We would like to thank B. Eden, J. Maldacena, G. Korchemsky and A. Tseytlin for useful discussions. The work of the authors is supported by ERC STG grant 306260. L.F.A. is a Wolfson Royal Society Merit Award holder. 

\appendix

\section{Asymptotic expansions of available results}

The universal structure constants were computed up to three-loops in \cite{Eden:2012rr}. The final result takes the form

\begin{eqnarray}
\label{edenresult}
 N(j)  &= &2 \left(\frac{\Gamma(j+1+\frac{\gamma}{2})^2}{\Gamma(2j+1+\gamma)} -\frac{1}{4} \frac{\Gamma(j+1)^2}{\Gamma(2j+1)} \sum_{i=2}^\infty \zeta_i b_i \right) \times \\
&  &\times \left( 1+ a c_{1,2} +a^2 \left( \zeta_3 c_{2,1}+c_{2,4} \right) +a^3 \left( \zeta_5 c_{3,1} +\zeta_3 c_{3,3}+c_{3,6} \right) \right)
\end{eqnarray}
where 

\begin{equation}
a= \frac{g^2 N}{4\pi^2},~~~~~~b_2 = -\gamma^2 +\left( S_1(2j) -S_1(j) \right) \gamma^3+...,~~~b_3 =\gamma^3+...
\end{equation}
and the anomalous dimension, which depends on the spin, has an expansion $\gamma = \gamma_1  a+  \gamma_2  a^2 +  \gamma_3 a^3+...$. The anomalous dimension, as well as the $c_{i,j}$ are written in terms of Harmonic sums, for instance:

\begin{eqnarray}
\gamma_1 &=& 2 S_1\\
\gamma_2 &=& -S_{-3} -2 S_{-2} S_1-2 S_1 S_2 -S_3+2 S_{-2,1} \\
c_{1,2} &=& -S_2 \\
c_{2,1} &=& 3 S_1\\
c_{2,4} &=& \frac{5}{2} S_{-4}+S_{-2}^2+2 S_{-3} S_1 +S_{-2} S_2+S_2^2 +2S_1S_3+\frac{5}{2} S_4-2S_{-3,1}-S_{-2,2}-2S_{1,3}
\end{eqnarray}
and so on. The rest of the expressions can be found in \cite{Eden:2012rr} and the argument of all harmonic sums is $j$. Harmonic sums are defined by \cite{Harmonic}

\begin{eqnarray}
S_a(j) = \sum_{m=1}^j \frac{1}{m^a},~~~~~S_{a,b,c,...}(j) = \sum_{m=1}^j \frac{1}{m^a} S_{b,c,...}(m) \\
S_{-a}(j) = \sum_{m=1}^j \frac{(-1)^m}{m^a},~~~~~S_{-a,b,c,...}(j) = \sum_{m=1}^j \frac{(-1)^m}{m^a} S_{b,c,...}(m) 
\end{eqnarray}
We will be interested in the large $j$ behavior of the above expressions. For instance, it is easy to check

\begin{eqnarray}
S_1(j) &=& \log j +\gamma_e+...,~~~~~S_{-1}(a) = -\log 2+...\\
S_a(j) &=& \zeta_a+...,~~~~~~~~~~~~~~~S_{-a}(j) =-\frac{2^{a-1}-1}{2^{a-1}} \zeta_a+... ~~~~a>1
\end{eqnarray}
Furthermore, harmonic numbers with the first entry different from one are finite in the large $j$ limit. The anomalous dimension of twist-two operators is known to several loops and their large $j$ expansion has been computed for instance in \cite{Kotikov:2004er}
\begin{eqnarray}
\gamma_1 &=& 2 \left( \log j+ \gamma_e \right) + {\cal O}(1/j)\\
\gamma_2 &=&-\zeta_2  \left( \log j+ \gamma_e \right) -\frac{3}{2} \zeta_3 + {\cal O}(1/j)\\
\gamma_3 &=& \frac{11}{4} \zeta_4  \left( \log j+ \gamma_e \right) +\frac{1}{2} \zeta_2 \zeta_3 +\frac{5}{2} \zeta_5+ {\cal O}(1/j)
\end{eqnarray}
Furthermore, we have computed the asymptotic expansions for the functions $c_{i,j}$ appearing in (\ref{edenresult}) and have obtained \footnote{The constant piece in $c_{3,6}$, as well as the others, can be computed with standard packages,for instance \cite{HarmonicSums}, but its form is quite involved.}

\begin{eqnarray}
c_{1,2} &=& -\zeta_2 + {\cal O}(1/j)\\
c_{2,1} &=& 3 \left( \log j+ \gamma_e \right) + {\cal O}(1/j)\\
c_{2,4}&=& -\frac{3}{2} \zeta_3  \left( \log j+ \gamma_e \right) +5 \zeta_4 + {\cal O}(1/j) \\
c_{3,1}&=& -\frac{25}{2} \left( \log j+ \gamma_e \right) + {\cal O}(1/j)\\
c_{3,3}&=& \frac{4}{3}  \left( \log j+ \gamma_e \right)^3 -\zeta_2  \left( \log j+ \gamma_e \right) -\frac{17}{6} \zeta_3 + {\cal O}(1/j) \\
c_{3,6}&=&  \frac{4}{3} \zeta_3  \left( \log j+ \gamma_e \right)^3 +(10\zeta_5-\zeta_2 \zeta_3)  \left( \log j+ \gamma_e \right)+ \frac{17}{12}\zeta_3^2 - \frac{591}{140}\zeta_2^3+ {\cal O}(1/j)
\end{eqnarray}
This gives the following expansion for $N(j)$, up to terms of order ${\cal O}(1/j)$:

\begin{eqnarray*}
&\log \left(\frac{N(j)}{N(j)_{tree}} \right) = -\left( 2\log 2 \log j'+\zeta_2 \right) a + \left(\zeta_2  \log^2 j'  + (\zeta_2 \log 2 +\frac{3}{2} \zeta_3 )  \log j'+\frac{3}{2} \zeta_2^2 +\frac{3}{2} \log 2 \zeta_3 \right) a^2+\\
&+ \left( \frac{2}{3} \zeta_3 \log^3 j' -\zeta_2^2 \log^2 j' + \left(-\frac{5}{2} \zeta_5 - 2\zeta_2 \zeta_3 -(\zeta_2^2+\frac{1}{4}\zeta_4) \log 2 \right) \log j'  +{\cal O}(1) \right) a^3+...
\end{eqnarray*}
where $j'=j e^{\gamma_e}$. The coefficient in front of each power of $\log j'$ can be seen to be exactly reproduced from our proposal (\ref{Cproposal}). Furthermore, the two results precisely agree if we choose

\begin{equation}
F(a) = -\frac{1}{2} \zeta_2 a+ \frac{7}{8} \zeta_2^2 a^2 +\left( \frac{\zeta_6}{128}-\frac{13}{8}\zeta_2^3-\frac{1}{8}\zeta_2 \zeta_4 \right) a^3+...
\end{equation}


\end{document}